\documentclass[pre,amsmath,amssymb,twocolumn,superscriptaddress,showpacs]{revtex4}

\usepackage{amsmath,amssymb}
\usepackage[usenames]{color}
\usepackage{amssymb}
\usepackage{grffile}
\usepackage[pdftex]{graphicx}
\usepackage{amsmath, amstext, amssymb, amsfonts, amsxtra}
\usepackage{textcomp}
\usepackage{xspace}

\def \ell{{d}}

\newcommand{\sutd}{Singapore University of Technology and Design, 20 Dover Drive, 138682 Singapore}

\begin{document}

\title{Work and Efficiency of Quantum Otto Cycles in Power Law Trapping Potentials}                   

\author{Yuanjian Zheng} 
\affiliation{\sutd}
\author{Dario Poletti}
\affiliation{\sutd}

\begin{abstract} 

We study the performance of a quantum Otto cycle driven by trapping potentials of the form $V_t(x) \sim x^{2q}$. This family of potentials possesses a simple scaling property which allows for analytical insights into the efficiency and work output of the cycle. We show that, while both the mean work output and the efficiency of two Otto cycles in different trapping potentials can be made equal, the work probability distribution will still be strongly affected by the difference in structure of the energy levels. Lastly, we perform a comparison of quantum Otto cycles in various physically relevant scenarios and find that in certain instances, the efficiency of the cycle is greater when using potentials with larger values of $q$, while, in other cases, with harmonic traps.  
\end{abstract}

\maketitle {}

The study of thermodynamics at the nanoscale has been the subject of intense interest in recent years (for some recent reviews see \cite{CampisiTalkner2011,Seifert2012}). As with classical thermodynamics, one of the main subject of study is the performance of heat engines. At small scales, heat engines are bound to produce not a deterministic but rather a probabilistic work output because of the relative importance of thermal and quantum fluctuations. In recent years, the non-equilibrium work fluctuations have been related to the change of free energy at equilibrium by the Jarzynski's equality \cite{Jarzynski1997}. This result can now be understood as a consequence of the Crooks' equation \cite{Crooks1998}, under the overarching category of fluctuation theorems. \cite{CampisiTalkner2011}. 

Important recent developments of the study of heat engines include the possibility, both in classical and quantum systems, to externally drive a system such that a physical process is adiabatic despite being executed in a finite time \cite{DemirplakRice2003, DemirplakRice2005, Berry2009, ChenMuga2010, TorronteguiMuga2013, DelCampo2013, DengGong2013, Jarzynski2013, DelCampoPaternostro2013, DeffnerDelCampo2014, SchaffLabeyrie2008, SchaffLabeyrie2010, BasonMorsch2012}. This `counteradiabatic' driving protocols would allow the possibility of achieving highly efficient adiabatic-like engines with finite power.    

Thus far, quantum thermodynamic cycles and processes have focused primarily on harmonic systems \cite {AbahLutz2012,RossnagelLutz2014, AbahLutz2014, QuanNori2007, Quan2009, TalknerHanggi2008, GalveLutz2009, Horowitz2012, FordBinnie2012}, with study of anharmonic potentials limited to weak first order perturbations of a frequency modulated harmonic oscillator \cite{DeffnerLutz2010} and the experimental verification of non-Gaussian behavior \cite{BickleBechinger2006}. Heat engine cycles using anharmonic traps present very different energy level spacing from harmonic ones thus resulting in very different work probability distributions. One interesting family of traps is that of even power law potentials which are proportional to $x^{2q}$, where $q$ is a positive integer number and $x$ is the position coordinate of the system. The scaling property of these potentials has been used in \cite{Jarzynski2013,DeffnerDelCampo2014} to engineer counteradiabatic driving protocols.   
In this paper we investigate further this class of trapping potentials, which includes both harmonic and anharmonic functions, focusing on utilizing the geometry of the trapping potential to tame the work fluctuations in a quantum heat engine. We examine cycles for which the average work output and efficiency are the same, and also scenarios in which the maximal and minimal temperatures of the quantum gas in the cycle are fixed. We show that, not only the work fluctuations are different for different $q$, but that, depending on the comparative scenarios analyzed, either the harmonic or anharmonic potentials can provide larger efficiency and/or work output. 

\begin{figure}
\centerline{\includegraphics[width=\columnwidth]{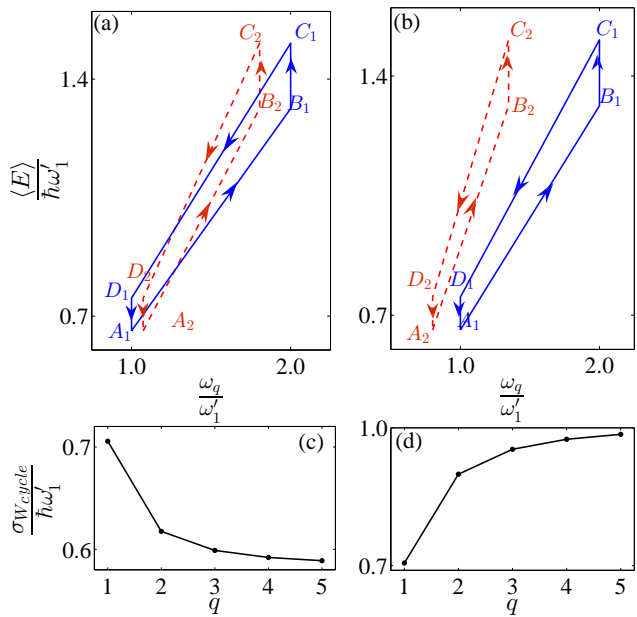}}
\caption{(Color online) (a,b) Mean energy $\langle E\rangle$ vs. trapping parameter $\omega_q$ for a quantum Otto cycle. The continuous blue lines represent the cycle in a harmonic trap ($q=1$), while the dashed red lines represent a cycle in an anharmonic trap with $q=2$. (c,d) Standard deviation of the work distribution $\sigma_{W_{cycle}}$ vs. the anharmonic parameter $q$. In all the plots, the mean energy at each vertex of the cycle is the same (`matched energies condition'). In (a,c) the initial temperature $1/\beta_{A_q}$ is the same for the two different confining potentials, while in (b,d) the initial volume, $V_{A_q}$, is matched. The extremal temperatures for the cycle in a harmonic trap are $1/\beta_{A_1}=1/2\;(\hbar\omega_1')^{-1}$ and $1/\beta_{C_1}=5/4\;(\hbar\omega_1')^{-1}$. } \label{fig:cycle} 
\end{figure}

The systems we study obey the Hamiltonian 
\begin{equation} \label{eq:hamiltonian}
\hat{H}(q,\omega_q)= -\frac{\hbar^2}{2m}\frac{\partial^2}{\partial x^2}+ \frac{1}{2}m \left(\omega_q\;\hat{x} \right)^{2q} 
\end{equation}
where $q=1,2,3...$ and $\omega_q$ represents the magnitude of the generalized trapping potential \cite{Jarzynski2013,DeffnerDelCampo2014}. Note that this generalized trapping potential reverts to the harmonic oscillator for $q=1$ and to the `infinite' box potential for $q=\infty$. 
The system undergoes a cycle consisting of four processes [see Figs.\ref{fig:cycle}(a,b) for a schematic representation]: First an adiabatic compression from a state $A_q$ ($q$ is the anharmonic parameter mentioned earlier), characterized by $\omega_q=\omega_q'$, to $B_q$, where the amplitude of the trapping is $\omega_q''>\omega_q'$ (no coupling to any thermal bath during this process); This is followed by a heat exchange at constant Hamiltonian parameters (no work is done nor received) from $B_q$ to $C_q$ due to a weak coupling to a thermal reservoir; The third process is an adiabatic expansion from $C_q$ to $D_q$  (no coupling to any thermal bath in this process either); Lastly, a heat exchange with the cold reservoir at constant Hamiltonian parameter brings the system back to state $A_q$. The cycle is fully determined by the choice of $\omega_q'$, $\omega_q''$ and by the temperatures $1/\beta_{A_q}$ and $1/\beta_{C_q}$ (we will refer to these last two parameters as the `extremal' temperatures because they are the lowest and highest temperatures achieved in the system). 

During the thermodynamic cycle, energy is exchanged under the form of work and heat transfer. When the Hamiltonian parameter $\omega_q$ varies from a value $\omega_q'$ to $\omega_q''$ following a process $p$, the (inclusive) work is described by the work probability distribution function $P(W_p)=\sum_{m,n}\delta(W_p-E_m''+E_n')\mathcal{P}^{m,n}P_n$ where $E_m''$ and $E_n'$ are the eigenvalues of, respectively $\hat{H}(q,\omega_q')$ and $\hat{H}(q,\omega_q'')$, $P_n$ is the initial thermal probability of occupation of the energy eigenvalue $n$ and $\mathcal{P}^{m,n}$ is the transition probability from the energy eigenstate $n$ to the eigenstate $m$ relative to the process $p$ (see for example \cite{CampisiTalkner2011}). When the system instead undergoes solely a heat exchange, for example between states $B_q$ and $C_q$, the heat transferred can be computed as the difference of the mean energies $\langle Q_{B_q \rightarrow C_q}\rangle=\langle E \rangle_{C_q}-\langle E \rangle_{B_q}$. And the efficiency of the cycle, $\eta_q$, is defined, as per usual, by the ratio of the modulus of net work done divided by the heat transferred into the system $\eta_q=-\left(\langle W_{A_q \rightarrow B_q}\rangle + \langle W_{C_q \rightarrow D_q}\rangle\right)/\langle Q_{B_q \rightarrow C_q}\rangle$ \cite{coupling}. 

While the cycle we study is commonly known as the Otto cycle, it should be noted that in the `classical' Otto cycle, a process in which no work is done or received corresponds to an isochoric process (no change in volume), but in the systems and regimes analyzed here, a process with no work transfer is obtained when the parameters of the Hamiltonian are kept unchanged. In this process, the volume occupied by the gas $V=\sqrt{\langle \hat{x}^2\rangle}$ does change, as heat is introduced into the system \cite{cstvolume}. 


To gain a deeper insight into the problem we rescale the Hamiltonian using the dimensionless coordinate $X=\left(\frac{m}{\hbar}\right)^\frac{1}{1+q} (\omega_q)^{\frac q{1+q}}x$. The dimensionless Hamiltonian $\hat{\mathcal{H}}_q$ is thus 
\begin{equation} 
\hat{\mathcal{H}}_q=\frac{\hat{H}(q,\omega_q)}{m^{\frac{1-q}{1+q}}(\hbar\omega_q)^{\frac{2q}{1+q}}}= -\frac{1}{2}\frac{\partial^2}{\partial X^2}+ \frac{1}{2}\hat{X}^{2q} \label{eq:dimensionless_hamiltonian}  
\end{equation}
with eigenvalues $e_{n,q}$. 
From \eqref{eq:dimensionless_hamiltonian}, we observe that the $n^{th}$ energy eigenvalue of $\hat{H}$, $\varepsilon_{n} (\omega_q)$, can be written as $\varepsilon_{n}=m^\frac{1-q}{1+q}\left(\hbar\omega_q\right)^\frac{2q}{1+q} e_{n,q}$. 

Now, considering a process in which the Hamiltonian parameter of the trapping potential is changed from $\omega_q'$ to $\omega_q''$, we can state, using the aforementioned scaling argument, that the ratio between two energy levels of the same order, will only depend on the ratio of parameters of the trapping potential, namely 
\begin{equation} 
\mu_q \equiv \frac{\varepsilon_{n}(\omega_q')}{\varepsilon_{n}(\omega_q'')} = \left( \frac{\omega_q'}{\omega_q''} \right)^\frac{2q}{1+q} \label{eq:mu}
\end{equation}
In the following we will refer to $\mu_q$ as the energy ratio parameter. For a given pair of parameters such that $\omega''_q > \omega'_q$ (compression), the energy ratio parameter is bounded: $0<\mu_{q}<1$. 
Moreover, the scaling properties of the Hamiltonian (\ref{eq:hamiltonian}) implies that states are always thermal during adiabatic processes. In fact, for any adiabatic process, where the population of each energy level remains unchanged, an initial thermal state remains thermal so long as $\beta \varepsilon_n(\omega_q')=\beta' \varepsilon_n'(\omega_q'')$ (where $\beta'$ has to be the same for every state $n$) \cite{Quan2009}. Using Eq.(\ref{eq:mu}) we can then derive the inverse temperature of the final state $\beta'=\beta\varepsilon_n(\omega_q')/\varepsilon_n(\omega_q'')=\mu_q\beta$, which is independent of $n$.

Using the energy ratio parameter $\mu_q$, it is also possible to express the average work of an adiabatic process in a simple form. Considering for example the compression from $A_q$ to $B_q$, the average work is written as    
\begin{align}
\langle W_{A_q\rightarrow B_q}\rangle &=\sum\limits_{m,n}\left[\varepsilon_m(\omega_q'')-\varepsilon_{n}(\omega_q')\right]\mathcal{P}^{mn} P_n^{0} \nonumber 
\\ &=\left(\frac{1}{\mu_q}-1\right)\sum \limits_{n} \varepsilon_n(\omega_q')\frac{e^{-{\beta\varepsilon_n(\omega_q')}}}{\sum \limits_{l}e^{-\beta_{A_q}\varepsilon_{l}(\omega_q')}} \nonumber  
\\ &=\frac{1-\mu_q}{\mu_q}\langle E_{A_q}\rangle \label{eq:adicomp}
\end{align}
where $\langle E_{A_q}\rangle$ is the average energy of the initial state ${A_q}$ \cite{adiabatic}. 
Furthermore, it is also possible to write the efficiency of the cycle in terms of the energy ratio parameter.  
\begin{align} 
\eta_q & = 1-\mu_q \label{eq:eff}
\end{align}
where, in reference to Eq.(\ref{eq:adicomp}), we have used $\langle W_{A_q\rightarrow B_q}\rangle=(1/\mu_q-1)\langle E_{A_q}\rangle$, $\langle W_{C_q\rightarrow D_q}\rangle=(\mu_q-1)\langle E_{C_q}\rangle$ and $\langle Q_{B_q\rightarrow C_q} \rangle=\langle E_{C_q}\rangle-\langle E_{A_q}\rangle/\mu_q$. 
In classical thermodynamics the efficiency of the Otto cycle can be written as a function of only the compression ratio $\kappa_q=V_{A_q}/V_{B_q}$ (the ratio between the volume before and after the compression), and of the adiabatic parameter $\gamma=C_p/C_v$ (the ratio between the heat capacity at constant pressure and at constant volume). In this spirit, we write $\mu_q$ as a function of the ratio of volumes of the quantum gas by making use of the rescaling of $X$ and the definition of volume $V$    
\begin{equation} 
\kappa_q=\frac{V_{A_q}}{V_{B_q}}= \left( \frac{\omega_q''}{\omega_q'} \right)^\frac{q}{1+q}=\frac 1 {\sqrt{\mu_q}} \label{eq:compratio}
\end{equation}
This allows us to write the efficiency of the Otto cycle as      
\begin{equation} 
\eta_q=1-\frac 1 {\kappa_q^{\gamma-1}} = 1 - \frac{\beta_{B_q}}{\beta_{A_q}} \label{eq:effcomp}
\end{equation}
which is the same expression as the efficiency of a classical Otto cycle \cite{QuanNori2007} since, for our family of potentials, $\gamma=3$ \cite{gamma}. 

Eq.(\ref{eq:effcomp}) shows that, independent of the value of $q$, the efficiency of the cycle is only a function of the ratio of temperature at the ends of the compression process. It is thus possible to either compare the efficiency of engine cycles with different values of $q$ or adjust their parameters such that the engine cycles would have the same efficiency. It is also possible to adjust the parameters of engine cycles with different $q$ such that the mean energy at each vertex of the cycle are identical (we will refer to this as `matched energies condition'). This requirement will force both cycles not only to have the same efficiency, but also the same average transfers of heat and work in each process. While the `matched energies condition' guarantees that mean values obtained for the engine cycles are the same, the work fluctuations are bound to be different, due to the different energy level structure of traps with different $q$.  

In Figs.\ref{fig:cycle}(a,b) we represent the Otto cycle in an average energy, $\langle E\rangle$, against trapping parameter, $\omega_q$, diagram. The blue continuous line represents a cycle for $q=1$ while the red dashed line is used for $q=2$. The parameters $\omega_q'$, $\omega_q''$, $\beta_{A_q}$ and $\beta_{C_q}$ have been chosen such as to fulfill the `matched energies condition'. Note however, that the `matched energies condition' does not uniquely define all the parameters. In order to do so, we add another, physically relevant, condition: In Figs.\ref{fig:cycle}(a,c) we have chosen the parameters such that the initial volume $V_{A_q}$ is the same for all $q$, while in Figs.\ref{fig:cycle}(b,d) the initial temperatures $1/\beta_{A_q}$ are the same for all $q$. The standard deviation of the work fluctuations, $\sigma_{W_{cycle}}=\sqrt{\langle W_{cycle}^2 \rangle - \langle W_{cycle} \rangle^2}$, as $q$ varies, is shown in Figs.\ref{fig:cycle}(c,d). 
Here the work probability distribution for a cycle is given by 
\begin{align} 
P(W_{cycle})=&\sum \delta(W_{cycle}-W_{A_q\rightarrow B_q}-W_{C_q\rightarrow D_q}) \times \nonumber \\ 
& P(W_{A_q\rightarrow B_q})P(W_{C_q\rightarrow D_q})
\end{align} 
where the summation includes all possible values of $W_{A_q\rightarrow B_q}$ and $W_{C_q\rightarrow D_q}$ \cite{coupling}. 
Figs.\ref{fig:cycle}(c,d) also clearly show that whether the work fluctuation actually grows or decreases with an increasing anharmonic parameter $q$, depends strongly on the additional matching condition (same initial temperature, volume or any other relevant physical quantity).  

%
\begin{figure}
\includegraphics[width=\columnwidth]{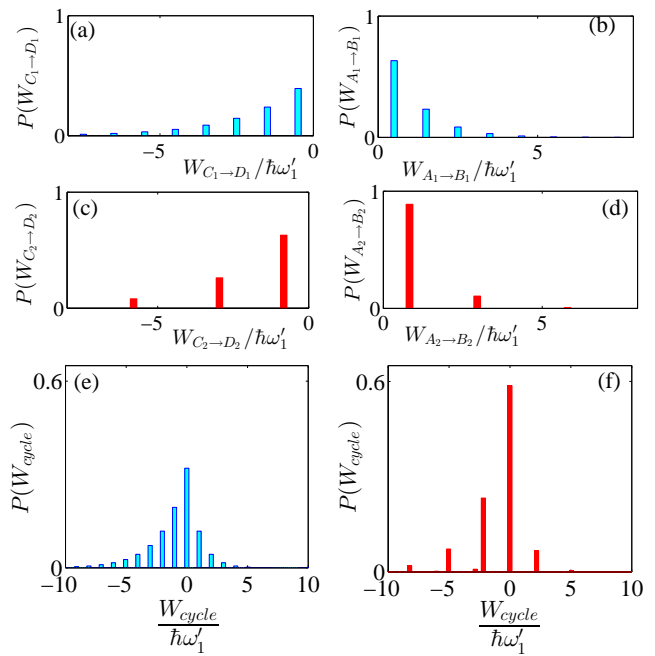}
\caption{(Color online) (a,b) Probability distribution of work, $P(W_p)$, for the adiabatic compression and expansion processes of the cycle in the harmonic trap (q=1). (c,d) $P(W_p)$ for the compression and expansion processes in the anharmonic trap with $q=2$. (e,f) Probability distribution of work for the full quantum Otto cycle for the harmonic (e) and anharmonic, $q=2$, trap (f). 
Cycle parameters (corresponding to 'matched energies condition' and also matching of the intial temperature $1/\beta_{A_q}$) are: $\omega_{1}''/\omega_{1}'=2$, $\omega_{2}'/\omega_{1}'=1.392$, $\omega_{2}''/\omega_{1}'=2.341$, $\beta_{A_1}=\hbar\omega_{1}'$, $\beta_{C_1}=1/4\;\hbar\omega_{1}' $.}\label{fig:workdistr}
\end{figure}
The different work fluctuations for various values of $q$ are rooted in the difference in work probability distributions of the various anharmonic parameters $q$ which is evident in Fig.\ref{fig:workdistr}. In particular, Figs.\ref{fig:workdistr}(a,c) show the work probability distribution for the expansion process, between $C_q$ and $D_q$. In Fig.\ref{fig:workdistr} we are using the `matched energies condition' and, to uniquely define the parameters of the cycles, we also chose the same temperature $1/\beta_{A_q}$ for both cycles (for the exact parameters, see the caption). Obviously, the histograms are not equidistant in Fig.\ref{fig:workdistr}(b) because, unlike in Fig.\ref{fig:workdistr}(a), the energy levels are not equidistant. It is also noticeable that the standard deviation in Fig.\ref{fig:workdistr}(c) is different from Fig.\ref{fig:workdistr}(a). These two aspects (the non equidistance of energy levels and the different variance) are also well represented in Figs.\ref{fig:workdistr}(b,d) where the net work probability distributions of the compression process for $q=1$ and $q=2$ are depicted respectively. 
In Figs.\ref{fig:workdistr}(e,f) we show instead the work probability distribution for the full cycle. We notice clearly that the average work output is negative and that the standard deviation is different in the two cases. We also note the presence of small histograms between much larger ones in Fig.\ref{fig:workdistr}(f) but not in Fig.\ref{fig:workdistr}(e). This is due to the presence of non equally-spaced energy levels for $q>1$. 

\begin{figure}
\centerline{\includegraphics[width=\columnwidth]{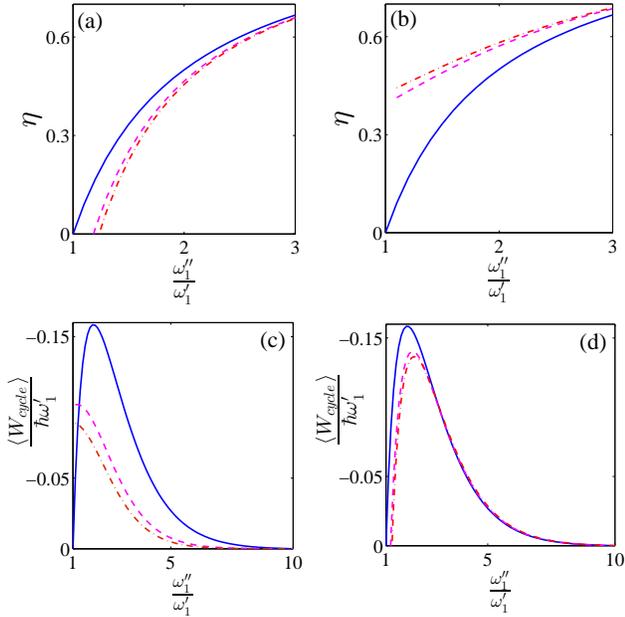}}
\caption{(Color online) (a,b) Efficiency of Otto cycles for different values of $q$ versus $\omega_1''$. (c,d) Comparison of average work done in a cycle $\langle W_{cycle} \rangle$ versus $\omega_1''$. In all the plots the continuous blue line is used for $q=1$, dashed pink line for $q=2$ and dot-dashed red line for $q=3$. The compared cycles have the same extremal temperatures ($\beta_{A_i}=\beta_{A_j}=10\;\hbar\omega'_1$ and $\beta_{C_i}=\beta_{C_j}=\hbar\omega'_1$ with $i,j=1,2,3$) and, for (a,c), matched extremal volumes ($V_{A_i}=V_{A_j}$ and $V_{C_i}=V_{C_j}$ with $i,j=1,2,3$), while in (b,d) matched extremal energies ($\langle E_{A_i}\rangle=\langle E_{A_j}\rangle$ and $\langle E_{C_i}\rangle=\langle E_{C_j}\rangle$ with $i,j=1,2,3$).}\label{fig:effVE}
\end{figure}

It is also important to investigate physically relevant scenarios that veer away from the stringent `matched energies condition'. For instance, the extremal temperatures $1/\beta_{A_q}$ and $1/\beta_{C_q}$ could be the same for two engine cycles (because the engines could be coupled to the same two thermal baths). To uniquely define all the parameters of the cycle, two more independent conditions are needed.  
In Figs.\ref{fig:effVE} two physically relevant examples are considered: (a,c) the case in which the extremal energies ($\langle E_{A_q}\rangle$ and $\langle E_{C_q}\rangle$) are also matched, and (b,d) the case in which the extremal volumes ($V_{A_q}$ and $V_{C_q}$) are identical.   
As shown in Figs.\ref{fig:effVE}(a,b), the efficiency of the engine cycles is an increasing function of $q$ when the extremal energies are the same, while it decreases with increasing $q$ when the extremal volumes are made the same. Furthermore, Figs.\ref{fig:effVE}(c,d) shows the net work output of the respective matching conditions, which may in fact, be either larger in the harmonic (continuous line) or in the matched anharmonic cases (lines with symbols) depending on the specific matching condition (extremal volumes or extremal mean energies etc.), and on the value of the matched $\omega_1''$. 

\begin{figure}
\centerline{\includegraphics[width=\columnwidth]{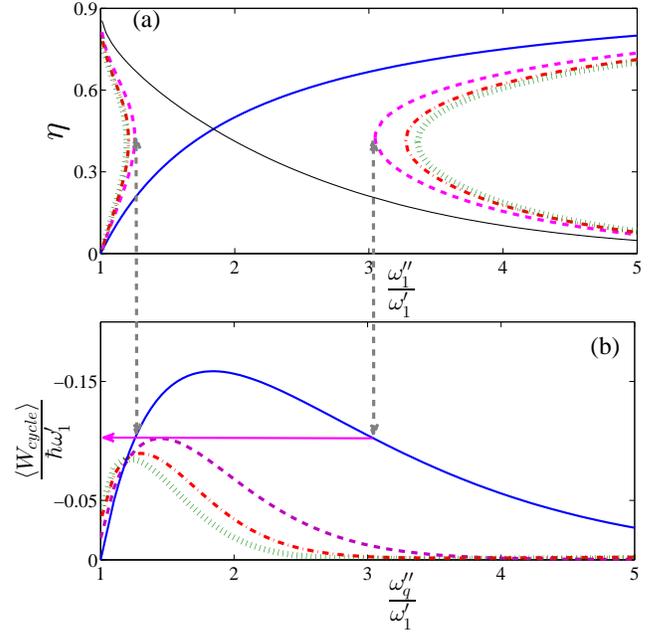}}
\caption{(Color online) (a) Efficiency of the cycle versus $\omega_1''$ for different anharmonic parameters $q$. The cycles compared have the same extremal temperatures ($\beta_{A_i}=\beta_{A_j}=10\;\hbar\omega'_1$ and $\beta_{C_i}=\beta_{C_j}=\hbar\omega'_1$ with $i,j=1,2,3,4$) and perform the same average work $\langle W_{cycle} \rangle$. Moreover, the initial mean energy $\langle E_{A_q} \rangle$ is the same. (b) Average work produced $\langle W_{cycle} \rangle$ versus $\omega_q''$ for same initial temperature $\beta_{A_q}$ and mean energy $\langle E_{A_q} \rangle$. In (a) and (b) the continuous blue and black lines are used for $q=1$, dashed pink line for $q=2$, dot-dashed red line for $q=3$ and dotted line for $q=4$. $\omega_{2}'/\omega_{1}'=0.9567, \omega_{3}'/\omega_{1}'=0.9137, \omega_{4}'/\omega_{1}' = 0.8802$. The pink arrow indicates the maximum work attainable for $q>1$ (for this matching condition), while the gray, dashed double arrows highlight the boundaries of the region for which engine cycles within a trap with $q>1$ cannot match the work of the harmonic case.} \label{fig:matchwork}   
\end{figure}

As a last case study, we choose the parameters $\omega_q'$ and $\omega_q''$ such that the average net work output and the mean initial energy $\langle E_{A_q} \rangle$ are matched to their corresponding values in the harmonic engine while keeping the extremal temperatures matched for engine cycles with different $q$. In this case we expect the efficiency to be dependent on the choice of $q$, which is clearly seen in Fig.\ref{fig:matchwork}(a), where each curve shows the efficiency for different values of $q$. We note that the figure consists of two separate curves for any given $q$ because the work output when the two extremal temperatures are fixed is not a monotonous function of $\omega_q''$, as can be seen in Fig.\ref{fig:matchwork}(b). This figure also shows that by fixing the extremal temperatures, there are values of work output which are attainable by the harmonic potential that cannot be generated by anharmonic potentials (to do so would require for instance, a much larger temperature at $C_q$ or a much lower temperature at $A_q$). It is for this exact reason that in Fig.\ref{fig:matchwork}(a) there exists a central region of the plot along the $\omega_1'' / \omega_1'$ axis for which there are no lines representing the efficiency of engine cycles within an anharmonic trap. 
In addition, Fig.\ref{fig:matchwork}(a) also shows that given a particular $\omega_q''$, the efficiency of all the anharnomic engine cycles (with the same average work and extremal temperatures) can either be always better (for low compressions) or worse (for large compressions) than that of the harmonic cycle. Note that the thin black continuous line in Fig.\ref{fig:matchwork} shows the efficiency of an engine cycle in a harmonic trap which still has the same $\langle E_{A_q} \rangle$ and $\langle W_{cycle} \rangle$ as the original solid harmonic curve, but is now achieved with a different $\omega_1''$. 
This analysis of Figs.\ref{fig:matchwork}(a,b) also teaches us that, while a desired amount of net work can be obtained with multiple $\omega_q''$, one particular choice of this value may give the best efficiency.    
     
In conclusion, we have studied a quantum Otto cycle driven by a particular class of trapping potentials. This family of potentials allows for the investigation of the relative performance of heat engines between harmonic and anharmonic configurations, which paves the way towards optimizing the work fluctuations by detailed design of the trapping geometry. In our analytical treatment that is made possible by the scaling properties of these potentials, we have found that, regardless of the values of the anharmonic parameter $q$, all engine cycles share the same expression for the efficiency, which corresponds to the classical expression. However, despite this apparent similarity in the expression for the efficiency, we have shown that the work probability distribution is still strongly affected even when both the average work output and efficiency of the cycles for different potentials are made identical. Subsequently, we have also analyzed cases in which engine cycles with different potentials are made to operate between the same extremal temperatures and studied various physically relevant scenarios for detailed and quantitative comparisons of the different engine cycles; we have found that, if the extremal energies of the cycle are matched, the engines within power law potentials with $q>1$ have greater efficiencies than those within a harmonic potential. On the contrary, if the extremal volumes are equal, then engines within harmonic potential are more efficient. Lastly, we have also shown that, for the case in which the extremal temperatures are the same for the two engines and the parameters are chosen such that the average work output is the same, then, for small compressions, engine cycles within an anharmonic potential are more efficient than cycles within harmonic trapping potentials, while for larger compressions the converse is true.  

An experimental realization of these engine cycles can be done with ultracold ions confined in a Paul trap, in which the trapping potential's shape is made anharmonic and changed periodically in time. 
This engine cycle could be produced following the procedure detailed in \cite{AbahLutz2012}. 

{\it Aknowledgments}: We are thankful to J. Gong and M. Mukherjee for fruitful discussions. We aknowledge support from the SUTD start-up grant (SRG-EPD-2012-045).



\begin{thebibliography}{10}
\bibitem{coupling} We have neglected, thanks to the weak coupling approximation, the work done or received by the system during the action of coupling the system itself to the bath.
\bibitem{cstvolume} As heat is introduced into the system with the Hamiltonian parameters held unchanged, work would have to be done on the system for the gas to occupy the same amount of volume.
\bibitem{adiabatic} We use the fact that, for an adiabatic process, $\mathcal{P}^{mn}=\delta_{mn}$ where $\delta_{m,n}$ is Kronecker's delta.
\bibitem{gamma} This is a consequence of the equation of state $PV=2\langle E\rangle$ which is valid for the family of trapping potentials considered.
\end{thebibliography}
\end{document}